# retinalysis-vascx: an explainable software toolbox for the extraction of retinal vascular biomarkers from color fundus images

Jose D. Vargas-Quiros[1,2], Michael J. Beyeler[5,6], Sofia Ortin-Vela[5,6], Sven Bergmann[5,6,7], Caroline C.W. Klaver[1,2,3,4] and Bart Liefers[1,2] on behalf of the VascX Research Consortium

**[1]Department of Ophthalmology, Erasmus University Medical Center, Rotterdam, the Netherlands**
**[2]Department of Epidemiology, Erasmus University Medical Center, Rotterdam, the Netherlands**
**[3]Department of Ophthalmology, Radboud University Medical Center, Nijmegen, the Netherlands**
**[4]Institute of Molecular and Clinical Ophthalmology, University of Basel, Switzerland**
**[5]Dept. of Computational Biology, University of Lausanne, Lausanne, Switzerland**
**[6]Swiss Institute of Bioinformatics, Lausanne, Switzerland**
**[7]Dept. of Integrative Biomedical Sciences, University of Cape Town, Cape Town, South Africa**

Correspondence:
Jose Vargas Quiros
Department of Ophthalmology
Erasmus Medical Center
Dr. Molewaterplein 40
Rotterdam 3015 GD
The Netherlands
Email: j.vargasquiros@erasmusmc.nl

Funding: This work was funded by the Swiss National Science Foundation grant no. CRSII5 209510.394

Word count: 8770

# ABSTRACT

**Purpose:** Automatic extraction of retinal vascular biomarkers from color fundus images (CFI) is crucial for large-scale studies of the retinal vasculature. We present VascX, an open-source software toolbox that extracts biomarkers from CFI artery-vein segmentations.

**Methods:** From a color fundus image, VascX computes artery and vein segmentation masks, extracts their centerlines, builds undirected and directed vessel graphs, and resolves vessel segments into longer vessels. A comprehensive set of biomarkers is derived, including vascular density, central retinal equivalents (CREs), and tortuosity. Spatially localized biomarkers may be calculated over grids placed relative to the fovea and optic disc.

**Results:** Our test-retest reproducibility analysis on repeat imaging of the same eye by different devices shows that most VascX biomarkers have moderate to excellent agreement (ICC > 0.5), with important differences in the level of robustness of different biomarkers. Our analyses of biomarker sensitivity to image perturbations and heuristic parameter values support these differences and further characterize VascX biomarkers.

**Conclusions:** VascX provides an explainable and flexible biomarker extraction toolbox that complements segmentation to produce reliable retinal vascular biomarkers. Our graph-based biomarker computation stages support reproducible, region-aware measurements suited for large-scale clinical and epidemiological research.

**Translational Relevance:** VascX supports oculomics research by enabling easy extraction of existing biomarkers and rapid experimentation with new biomarkers. Its robustness and computational efficiency facilitate scalable deployment in large databases. VascX is released via GitHub and PyPI with comprehensive documentation and examples to facilitate adoption by ophthalmic researchers and clinicians.

## Introduction

Artificial intelligence (AI) has rapidly changed ophthalmic research by enabling the quantitative, automated analysis of retinal imaging at scale. Modern deep learning (DL) applied to color fundus imaging (CFI) has demonstrated the ability to automatically recognize ophthalmic conditions with remarkable accuracy. In recent years, the field of oculomics has taken this a step further by utilizing the eye to understand systemic health[1–3]. Non-invasive CFI allows for time- and cost-efficient examination of the arteriolar and venular retinal vasculature, opening the door for large-scale analysis in existing population-based cohorts and clinical databases. Today, substantial evidence links key retinal vascular biomarkers to hypertension, kidney disease, stroke subtypes, and overall cardiovascular risk[2].

Retinal vascular biomarkers include the central retinal arteriolar/venular equivalents (CRAE/CRVE), artery-vein ratio (AVR), bifurcation angles, tortuosity/fractal dimension, and bifurcation counts. Such biomarkers are thought to capture changes in the vasculature due to hypertension, hemodynamic status, and network remodeling/efficiency, and have each been linked to target organ damage and incident vascular events[2,4].

Prior tools for automatically extracting retinal vascular biomarkers from CFIs have laid a strong foundation. Several systems made use of combined vessel segmentations (arteries and veins in a single segmentation map) to extract parameters such as fractal dimension (a measure of complexity), vessel density, vessel diameters, and tortuosity metrics[5,6]. Recently, datasets, models, and biomarker extraction pipelines have been developed to segment and analyze the artery and vein vessel trees separately [7–11]. PVBM provides a modular Python toolbox for computing a diverse set of vascular biomarkers from pre-segmented vessel maps, including branching angles, endpoints, and intersections[12]. AutoMorph delivers a comprehensive end-to-end DL pipeline that includes standardized morphological measurements like caliber, tortuosity, and measures of vascular complexity[9]. Together, these tools have made vascular analysis on CFIs more accessible to researchers

Nevertheless, important technical gaps remain. Most notably, PVBM and AutoMorph do not use the location of the fovea, and therefore do not support biomarker computation over regions defined relative to the optic-disc–fovea axis. The topological data representation in both tools is limited to an undirected graph, limiting the development of biomarkers that rely on the directedness of the vessel tree graph. Furthermore, both tools lack utilities for easily visualizing biomarkers on arbitrary CFIs, which limits interpretability.

Crucially, the cross-device reproducibility and the robustness of different retinal vascular parameters to imaging conditions have not been characterized before, making it difficult for researchers to judge their reliability, choose between biomarker variants, and understand the effect of configuration parameters or heuristic choices[13]. Although the test-retest reliability of biomarkers computed by AutoMorph pipeline was measured in previous work[14,15], this was done on images from a single fundus camera.

We address these technical and knowledge gaps by presenting and evaluating VascX, an open-source, well-documented, and easily modifiable Python biomarker extraction toolbox that utilizes artery-vein and optic disc segmentations and fovea locations for region-aware biomarker extraction[10]. VascX supports computation of localized biomarkers over regions defined relative to the fovea and optic disc landmarks, facilitating harmonized reporting and cross-device analyses. Biomarker visualizations can be generated on arbitrary input CFIs, facilitating explainability and human evaluation of the system.

Internally, VascX relies on a structured pipeline with four internal representations derived from artery and vein segmentation masks. Retinal vascular biomarkers are computed from the most appropriate representation stage (e.g., bifurcation angles computed from a directed graph, tortuosity computed over segments linked to an undirected graph). VascX is open source, well documented, and designed to facilitate biomarker innovation and large-scale, reproducible oculomics research.

We present an in-depth evaluation of VascX biomarker robustness via a cross-device test-retest reproducibility analysis, and analyses of biomarker sensitivity to simulated image perturbations and to changes in heuristic parameter values. Our test-retest analysis, in particular, makes use of extensive repeat CFI imaging from the Rotterdam Study cohort to test biomarker reproducibility across at least seven distinct CFI camera pairs, with differences in image quality, field of view, color and lighting characteristics, representative of large population and clinical datasets. Together, these analyses provide a comprehensive picture of retinal vascular biomarker reproducibility and inform their application in large-scale oculomics research.

# Methods

VascX was developed as an open-source software package for accurate and explainable biomarker computation that operates on image segmentations (usually generated by AI models) and outputs CSV file with biomarkers computed from them. The segmentation of arteries, veins, and the optic disc and the localization of the fovea have been addressed in previous work[8–10]. VascX processes these segmentations through four computation stages (skeletonization, undirected graph, directed graph, and resolved vessels, applied to arteries and veins separately. These stages are the foundation for the downstream computation of multiple biomarkers including vascular density, vascular calibers, central retinal equivalent (CRE), tortuosity, and others. Many of these biomarkers can be computed on specific regions of the CFI.  VascX was developed entirely in Python using standard data science and image manipulation packages: *numpy*, *opencv*, *Pillow*, *scikit-learn*, *skimage,* and other open-source dependencies. The pipeline can be executed in bulk on a folder containing suitable images (standard image formats such as PNG, JPG or DICOM) using the command line, or imported into Python code for advanced use cases. The following sections detail the computation stages, implemented biomarkers, region-aware computation, and intended usage of the VascX pipeline.

# Software Summary

VascX is available on GitHub[16] under an open-source license (GNU Affero General Public License v3.0). The package is also available in the Python package index (PyPI) under the name retinalysis-vascx. In this work, we present a first public version of the package. We will follow semantic versioning for future versions.

## Usage

The VascX pipeline can be installed as a Python package:

```
pip install retinalysis-vascx
```

We provide helpers and instructions for running in batches. The entire pipeline can be run in two stages:

**1. Segmentation**. To extract all the necessary segmentations from the images:

```
vascx run-models <INPUT_DIR> <SEGMENTATIONS_DIR>
```

This command will store model output segmentations and some intermediate files in a folder structure with matching filenames:

```
<SEGMENTATIONS_DIR>
  - preprocessed_rgb - preprocessed fundus images
  - artery_vein - artery-vein model segmentations
  - vessels - vessel model segmentations
  - disc - optic disc model segmentations
  - bounds.csv - contains the bounds of the fundus image
  - fovea.csv - model predictions of the fovea locations for each image
  - quality.csv - model estimations of CFI quality
```

This command makes use of the previously published VascX AI ensembles[10] to segment vessels, arteries, veins, and disc, obtain fovea locations, and estimate image quality. Note that this step requires a GPU with at least 8GB of memory. The folders (preprocessed_rgb, artery_vein, vessels, disc) contain matching filenames (one per input image). The CSV files contain one row per input image.

**2. Biomarker extraction**. The second stage operates on the output from the first stage:

```
vascx calc-biomarkers <SEGMENTATIONS_DIR> <OUTPUT_CSV_FILE> --feature_set full_v3 --n-jobs 8 --logfile <LOG_FILE>
```

Where <SEGMENTATIONS_DIR> is the path to the directory used in step 1, <OUTPUT_CSV_FILE> is the desired output file path, and <LOG_FILE> is an optional file path for logging errors and warnings. This step is CPU-only and benefits from parallel execution for performance. The *--n-jobs* option may be changed depending on the number of available CPU cores.

## Vascular biomarker computation

VascX processes the input artery-vein model segmentations into separate artery and vein masks. These are then processed through four core stages, each producing a different data representation:

- **Input masks**: `np.ndarray[bool]` per layer; optic disc and fovea metadata from segmentation models.
- **Stage 1 - Skeleton**: We first hole-fill the predicted binary vessel map to obtain a clean `binary` mask; when an optic disc mask is available, the disc region is excluded to prevent spurious centerlines. The skeleton is computed with `skimage` skeletonization on `binary`, yielding single-pixel-wide centerlines that preserve network topology. This pixel-level representation underpins coverage and diameter sampling and serves as the substrate for graph construction.
- **Stage 2 - Undirected graph**: From the skeleton, we build a NetworkX `Graph` whose nodes correspond to end/junction points and whose edges follow the chain of centerline pixels between them. Each edge carries a `Segment` that stores the ordered centerline, arclength and geometric summaries; splines are fitted on demand to the centerline to enable robust diameter and angle estimation. This representation supports segment-level filtering and aggregation without yet imposing flow direction.
- **Stage 3 - Directed digraph**: We orient the undirected graph into a `DiGraph` by rooting each connected component at the optic disc and directing edges away from the disc. Roots are identified as the end-node per component closest to the disc center. Vessel segments are directed edges carrying geometry and derived properties (length, median diameter, curvature). The directed graph representation is an intermediate step in the identification of bifurcations. A distance-transform–based assignment maps each pixel in the binary vessel mask (stage 1) to its nearest segment.

- **Stage 4 - Resolved vessels**: In previous stages, vessel segments are defined only between nodes in the skeleton/graph representation. This means that the presence of even a small vascular branch on a large vessel will cause it to be split into two segments. To facilitate the computation of biomarkers on longer, potentially more anatomically relevant vessel trajectories, we run a recursive vessel-resolution algorithm on the `DiGraph`. This reduces fragmentation from skeletonization while preserving topology for trajectory-level tortuosity and OD–fovea–aligned measures.

Finally, biomarkers are extracted from one or a combination of these data representations. Mask-based biomarkers, such as vascular density, are calculated on the input masks, while topological ones, such as bifurcation angles, use the directed graph and nodes extracted in stage 4. Figure 1 exemplifies the different stages of computation for a sample set of CFIs.

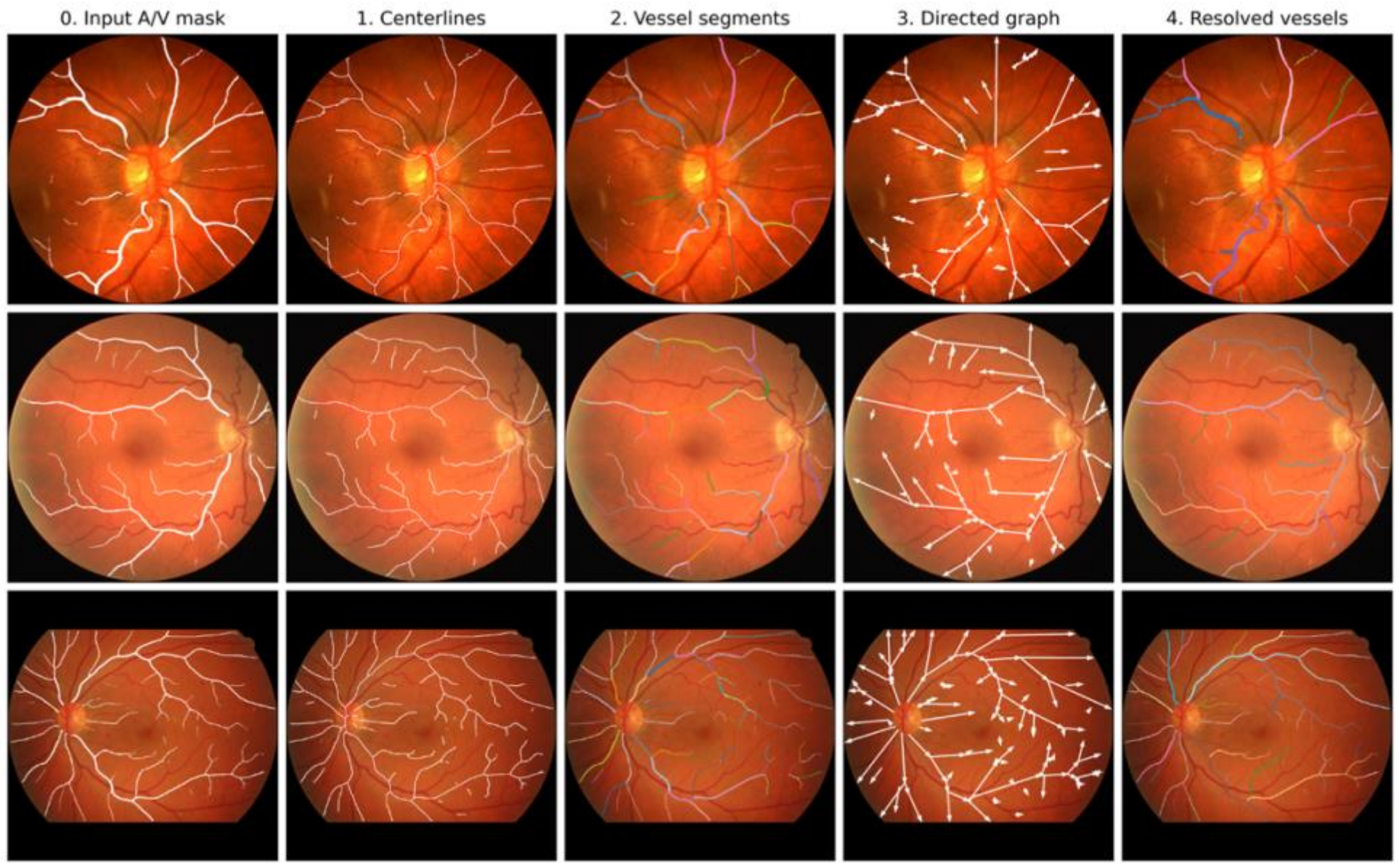

*Figure 1. From left to right, stages of computation in the VascX pipeline for three sample CFIs: 0) input binary mask for the arteries, 1) skeletonization or centerline computation, 2) vessel segments are extracted from the skeleton, and mapped back to their centerline and mask, 3) a directed graph consisting of multiple trees with the optic disc as root, 4) segments are resolved into potentially more anatomically meaningful structures for biomarker measurements.*

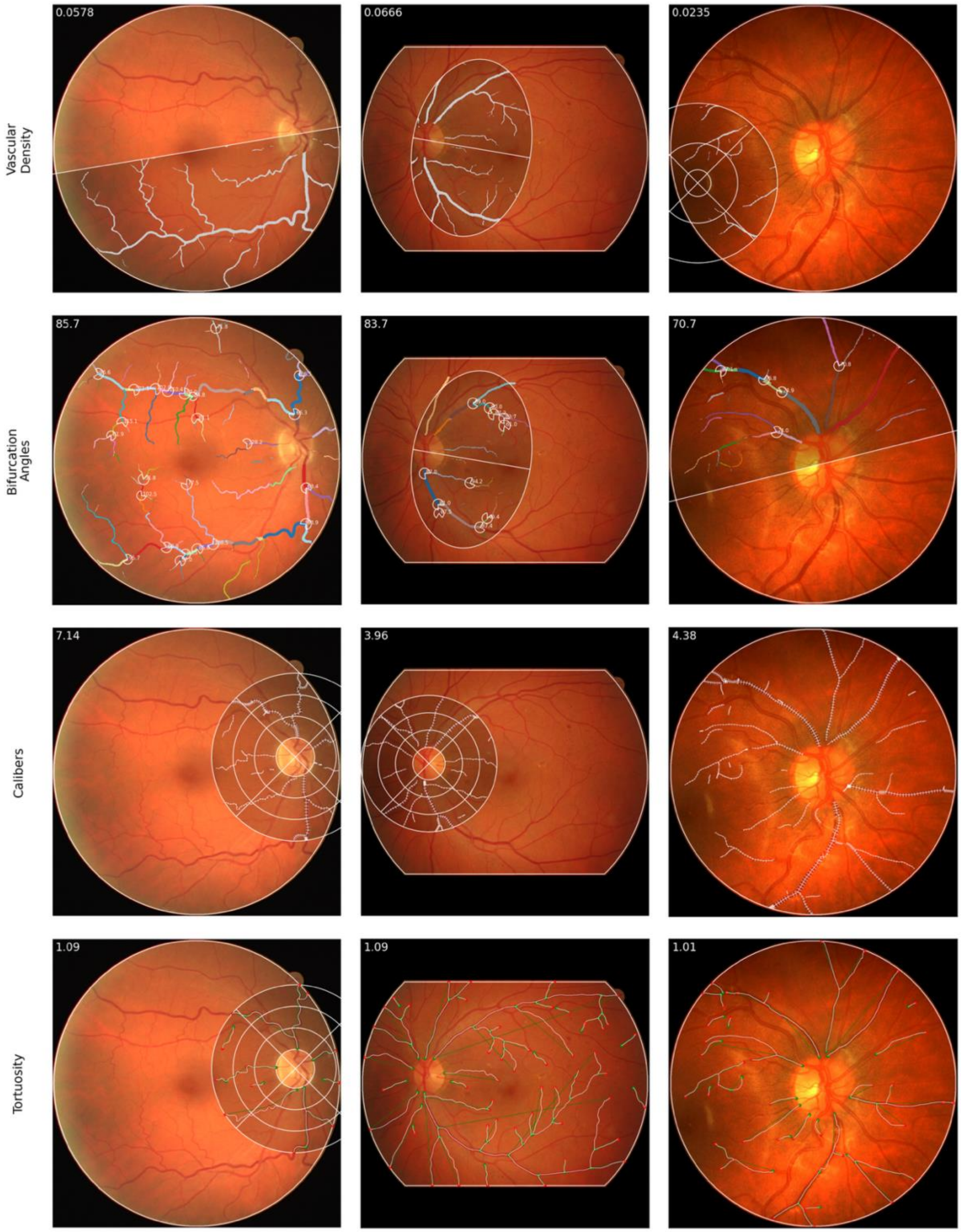

*Figure 2. Examples of VascX computation of Vascular Density, Bifurcation Angles, Caliber and Tortuosity biomarkers on fovea and optic disc-centered CFIs on different regions*

## Biomarkers

Below we describe the biomarkers currently implemented in VascX, with the exact quantity being estimated and the equations used. Throughout, B denotes the stage-1 binary vessel mask, S the set of eligible directed segments with lengths $\ell_i$, and R an analysis region of interest. Sample *explanations* of the biomarkers presented below – as generated by the pipeline – are shown in Figures 2 and 3, showing the measured vessels and landmarks and in the regions measured for region-local measurements

**VascularDensity.** The fraction of retinal area occupied by vessels in R, computed on the binary mask B:

$$D = \frac{|B \cap R|}{|R|}.$$

**BifurcationAngles.** For each bifurcation $b$ at position $p_b$, outgoing branch directions are estimated by sampling the branches' splines at distance $\delta$ from the node along each branch at points $q_1$ and $q_2$. Unit vectors $(u_b, v_b)$ are defined from the bifurcation point to the sample points:

$$u_b = \frac{q_1 - p_b}{\| q_1 - p_b \|}, \quad v_b = \frac{q_2 - p_b}{\| q_2 - p_b \|},$$

and the bifurcation angle is defined as the angle between these vectors:

$$\theta_b = \arccos(u_b \cdot v_b), \quad \theta_b \in [0°, 180°].$$

Angles exceeding 160° are discarded as non-bifurcating continuations. Summary statistics (e.g., mean/median) are reported across valid nodes.

**Caliber.** For each segment $i$, diameters are sampled along a spline fitted to its skeleton by projecting spline normals to the vessel boundary on B. The per-segment diameter is the median along its arclength. The reported caliber aggregates over eligible segments (length $\ell_i \geq \ell_{\min}$):

$$\text{Caliber} = g(\{d_i : i \in S\}),$$

where (g) is a robust statistic (median, or a vessel-length-weighted average).

**Tortuosity.** Three complementary measures are provided per segment (or per resolved vessel). Let $L_{\text{arc},i}$ be arclength and $L_{\text{chord},i}$ the end-to-end Euclidean distance.

- Distance factor:

$$T_i^{\text{DF}} = \frac{L_{\text{arc},i}}{L_{\text{chord},i}}.$$

- Curvature-based measure, using planar curvature $\kappa_i(s)$ and OD–fovea distance $d_{ODF}$ for scale normalization:

$$T_i^{\kappa} = \frac{1}{L_{\text{arc},i}} \int_0^{L_{\text{arc},i}} |\,\kappa_i(s)|\, ds \ \cdot \ d_{ODF}.$$

- Inflection count (number of curvature sign changes along the centerline):

$$T_i^{\text{INF}} = N_{\text{inflections}}^{(i)}.$$

When reporting a single score over multiple segments, length-weighted aggregation may be used for normalization:

$$T_{\text{tot}} = \sum_{i \in S} \left( \frac{\ell_i}{\sum_{j \in S} \ell_j} \right) t_i,$$

with $t_i$ any of the measures above.

**CRE (Central Retinal Equivalents).** CREs are calculated using the revised formulas by Knudtson et. al.[17] Concentric circles centered at the optic disc are intersected with the vessel network. At each radius $r$, up to $M$ vessel segments with the largest median diameters are retained. The vessel segments are paired and recursively reduced via the Hubbard rule with a constant $c$ (arteries: 0.88; veins: 0.95):

$$d \leftarrow c\sqrt{d_1^2 + d_2^2}$$

Where d1 and d2 are the diameters of the paired vessels. The formula is applied recursively until a single equivalent caliber $d_r$ remains. The final CRE is the median of $\{d_r\}$ across radii. For details about the pairing and reduction procedure please refer to Knudtson et. al. [17]

**TemporalAngle.** On each concentric circle of radius (r), the two dominant temporal vessels are identified by diameter and spatial continuity. The angle at the disc center is

$$\theta_r = \angle\big(\overline{OD\,p_1(r)}, \overline{OD\,p_2(r)}\big),$$

and the reported value is the median over radii.

**Sparsity.** Let $\mathrm{DT}(x)$ represent the distance transform over $R$, ie. the normalized Euclidean distance to the nearest vessel pixel (scaled by $d_{ODF}$). Over pixels in R we report either the mean or the largest local maximum:

$$S_{\text{mean}} = \frac{1}{|R|}\sum_{x\in R} \mathrm{DT}\,(x), \qquad S_{\text{max}} = \max_{x\in R\cap\text{local maxima}} \mathrm{DT}(x).$$

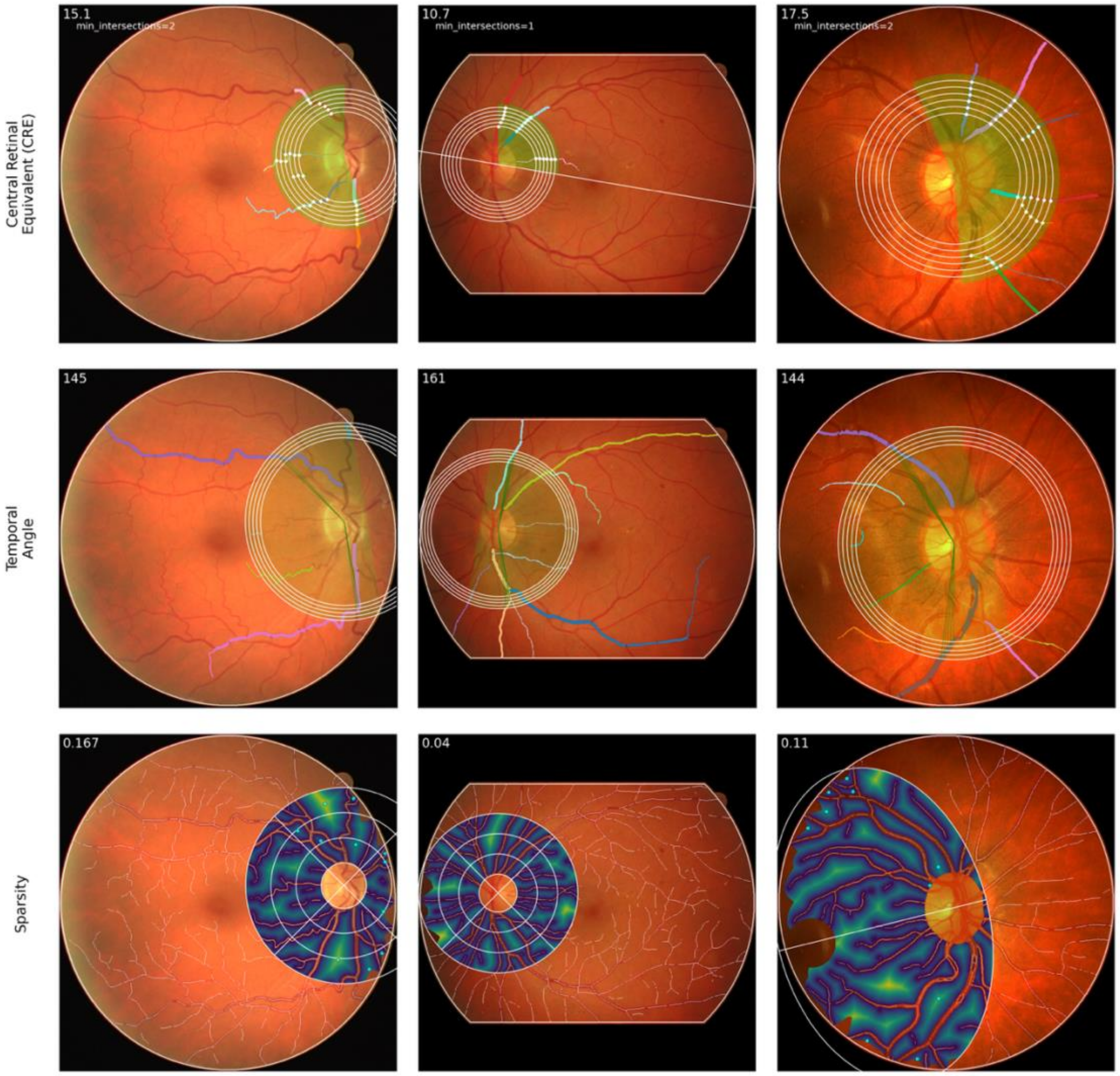

*Figure 3. Examples of VascX computation of Central Retinal Equivalent, Temporal Angle and Sparsity biomarkers on fovea and optic disc-centered CFIs on different regions.*

## Biomarker localization and region-awareness

Most biomarkers in VascX can be configured with an optional `roi` (region of interest) parameter that specifies the computation region. Regions are defined relative to a grid. A grid can be understood as a subdivision of the image space (eg. the ETDRS grid), and a region as a particular part of that subdivision (eg. the outer ring of the ETDRS grid, the superior hemifield, or the entire ETDRS grid). The optic disc mask and fovea position (e.g.,

OD–fovea axis) are usually used to scale and place the grid consistently relative to the retinal anatomy. VascX implements several standard and custom grids including the standard ETDRS grid for macula-centered measurements, flexible macula and disc-centered grids, and an elliptical grid designed to be visible on both macula and OD-centered images.

Depending on the region imaged and the field of view of the CFI, the region of interest may be out of bounds – not visible in the CFI. VascX can detect when the region of interest is (partially) out of bounds and return a null value rather than an invalid measurement. VascX models CFI bounds as a binary mask indicating the image region, computed using the retinalysis-fundusprep package[18], which fits a fundus circle and optional straight edges. For biomarkers configured with a region of interest, VascX calculates the fraction of the region that is within the CFI bounds. If the fraction is lower than a configurable threshold, biomarker computation returns a null value to indicate an invalid measurement. If the RoI lies sufficiently within image bounds, biomarker computation will proceed using only the structures that lie within it. Segments and full vessels are trimmed to the part of them that lies within the RoI. Nodes (endpoints, bifurcations) are included if their coordinates fall inside the region. Pixel/area-based measures (e.g., vascular density) are computed using the pixels inside the RoI mask only.

VascX includes the following predefined grids:

- `DiscCenteredGrid`: disc-anchored rings (`inner`, `center`, `outer`) and hemifields (`superior`, `inferior`, `nasal`, `temporal`, plus `left`/`right`), taking laterality into account. This grid is meant for consistent measurements on disc-centered images.

- `CircleGrid`: circle centered along the OD-fovea axis (radius derived from OD–fovea distance and disc size). This grid provides consistent measurements on macula-centered images. Regions: `FullGrid`, `Superior`, `Inferior`.

- `ETDRSGrid`: classic macula-centered 3mm grid with rings (`Center`, `Inner`, `Outer`), quadrants (`Superior`, `Inferior`, `Nasal`, `Temporal`, plus `Left/Right`), and subfields (`CSF`, `SIM`, `NIM`, `TIM`, `IIM`, `SOM`, `NOM`, `TOM`, `IOM`).

- `EllipseGrid`: ellipse centered midway between disc and fovea, major axis along OD–fovea. This grid has consistently high visibility in both optic disc and macula-centered images. Regions: `FullGrid`, `Superior`, `Inferior`.

- `HemifieldGrid`: superior/inferior half-planes split relative to the OD–fovea axis. Fields: `FullGrid`, `Superior`, `Inferior`. This grid is used to partition the CFI into superior and inferior hemifields relative to fovea and OD.

## Biomarker and grid scaling

While several VascX biomarkers such as tortuosity are unitless, others such as vessel calibers or CREs measure distances. VascX, by default, produces distance measurements in pixels and leaves the potential conversion of such measurements up to the user. If conversion factors are available, VascX supports an optional explicit mm_per_pixel scaling factor, provided per sample. When provided, this factor will be used for both conversion of output biomarkers into physical units and for scaling of grids and regions defined in physical units. Currently, the only grid in VascX with a scale defined in physical units rather than relative to the OD-fovea distance is the ETDRS grid, defined with a radius of 3mm per its original definition. However, when mm_per_pixel is not provided, VascX will scale the ETDRS grid using an assumed distance of 4.75mm between the optic disc center and the fovea to derive a mm/pix scaling factor – effectively resulting in scaling relative to the anatomy. The 4.75mm distance is a population-based average derived in previous work [19].

## Biomarker arguments

In addition to arguments that enable biomarker localization, VascX biomarkers have several optional arguments that affect computation. For example, the radius of the circles used for CRE computation or spline smoothing parameters used to compute tortuosity, can be configured for a particular biomarker. While the platform includes sensible defaults calibrated to work on most standard CFIs, the ability for researchers to easily customize biomarkers is a core feature of the platform. Supplementary Table 1 shows a summary of current biomarker arguments and briefly describes their role.

# Evaluation

We evaluated the test-retest reproducibility and sensitivity to image perturbations of retinalysis-vascx output biomarkers using diverse CF imaging data from the population-based Rotterdam Study[20].

## Evaluation database

The Rotterdam Study (RS) is a prospective population-based cohort study in Rotterdam, the Netherlands[20]. The RS includes repeated ophthalmic examinations across multiple study cohorts (RS-I to RS-IV). CF images have been collected since 1991 using multiple generations of CF cameras, including analog devices. Crucially, many RS visits involved repeat imaging of the same eye, captured within a few minutes of each other using different devices, enabling the analysis of test-retest biomarker reproducibility across devices.

## Test-retest biomarker reproducibility

We provide validation of the biomarkers from retinalysis-vascx via test-retest reproducibility experiments. We made use of CFI imaging from the RS to compare the agreement between biomarkers extracted from images of the same eye taken using different CF cameras during the same patient visit. The temporal overlap in the imaging devices used in the RS allowed us to analyze 8 device pairs resulting from overlap in imaging from 7 Topcon devices used to image the cohort between 1990 and 2026. The field of view, color, and imaging characteristics varied per device. Figure 4 shows sample images used in the evaluation for each device. We measured agreement via intra-class correlation coefficients (ICCs) using a Two-way Random Effects model – ICC(2,1), which assesses the absolute agreement of single measurements, treating both eyes and devices as random effects. We stratified the images by field (F1 – OD-centered and F2 – fovea-centered) and applied quality control using our sparsity measures. Biomarkers measured in pixel units (calibers and CREs) were transformed to a normalized scale by multiplying them by a scalar scaling factor calculated per device. The scaling factor was calculated between a pair of devices by fitting a Deming regression through the origin, and then computing canonical per-device factors multiplicatively, using the most common device (TOPCON TRC-50EX + DXC-950P) as an anchor. Figures 5 and 6 show the ICC for each *(biomarker, device)* pair along with the number of image pairs used in ICC computation. Note that a given pair was excluded from the score when one of the biomarkers was not computable on one of the two images (i.e., when *vascx* produced a *NULL* value). Table 1 shows the per-device means and limits of agreement (LoA) and Pearson r between paired samples for macula-centered (F1) images, for the five device pairs with the most samples.

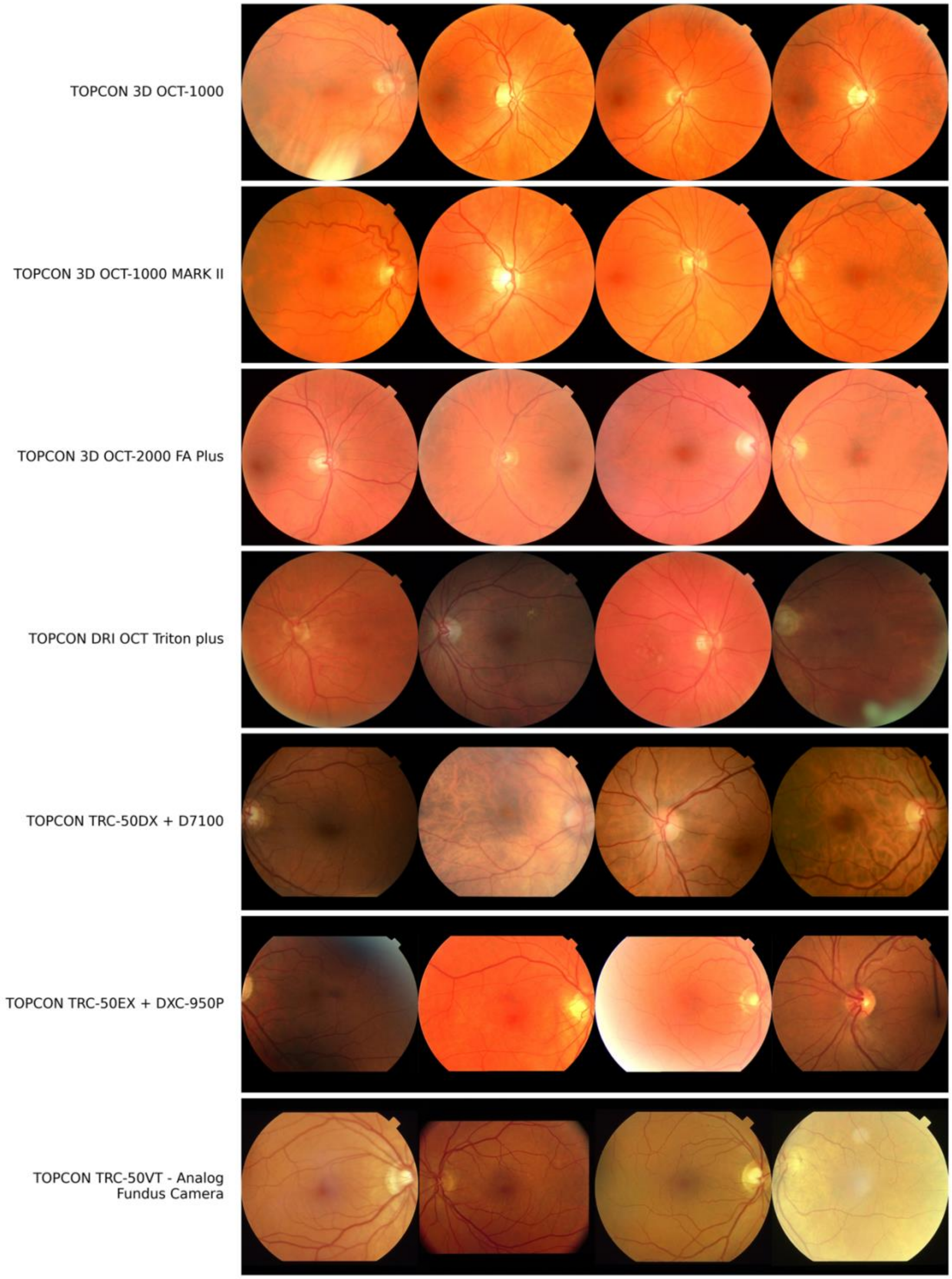

*Figure 4. Sample images from the devices used in the cross-device test-retest biomarker reproducibility experiments.*

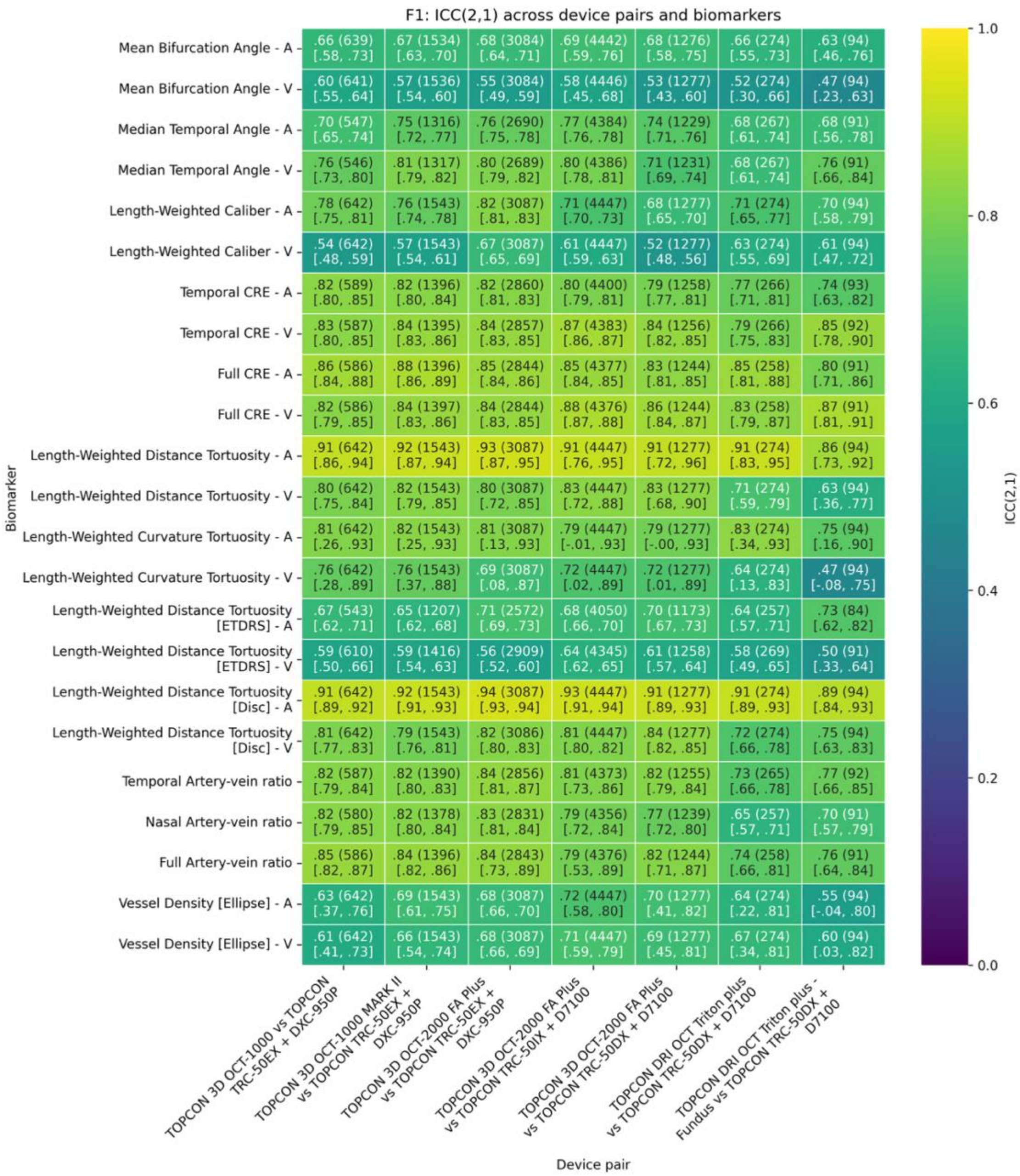

*Figure 5. Biomarker test-retest agreement between measurements from images of the same eye by different CFI devices, for field-1 (F1) images (optic disc centered). Intra-class correlation coefficients were computed for each device pair (X axis) with overlapping imaging (on the same eyes) in the Rotterdam Study. Each cell displays the ICC(2,1) score (two-way random effects, absolute agreement, single measures), the number of image pairs (N) used in the calculation of the score (in parenthesis) and the 95% confidence interval for the ICC.*

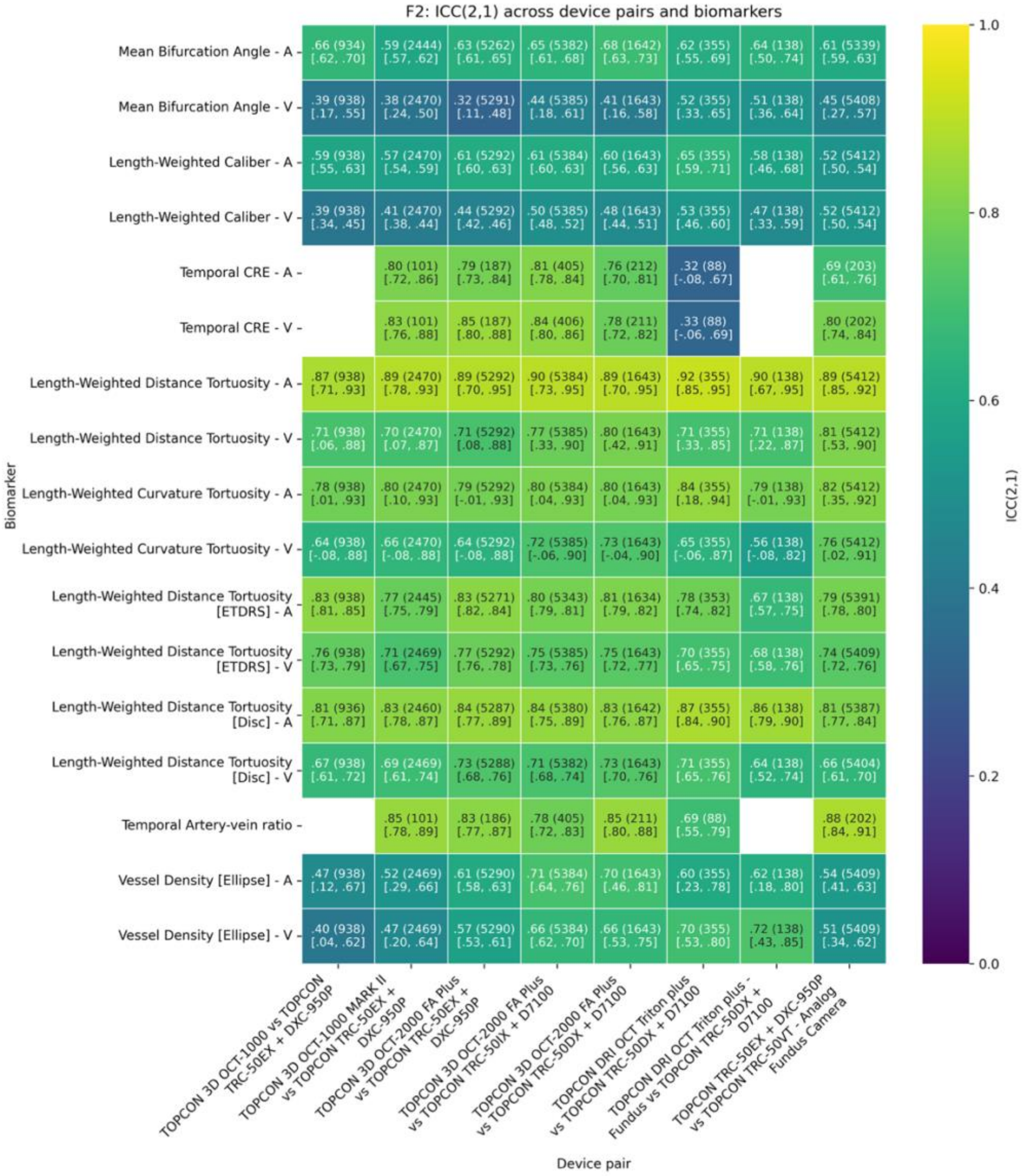

*Figure 6. Biomarker test-retest agreement between measurements from images of the same eye by different CFI devices, for field-2 (F2) images (macula-centered). Intra-class correlation coefficients were computed for each device pair (X axis) with overlapping imaging (on the same eyes) in the Rotterdam Study. Each cell displays the ICC(2,1) score (two-way random effects, absolute agreement, single measures), the number of image pairs (N) used in the calculation of the score (in parenthesis), and the 95% confidence interval for the ICC. Blank cells had fewer than 50 valid biomarker pairs and were excluded from computation.*

*Table 1. Biomarker means, limits of agreement (LoA), Pearson r and sample size (in parentheses) measured on pairs of images of the same eye by different CFI devices, for optic-disc-centered (F1) images. All device models refer to Topcon CFI devices used in the Rotterdam Study. LW stands for length-weighted.*

| **Biomarker** | **Statistic** | **3D OCT-1000 vs TRC-50EX + DXC-950P** | **3D OCT-1000 MARK II vs TRC-50EX + DXC-950P** | **3D OCT-2000 FA Plus vs TRC-50EX + DXC-950P** | **3D OCT-2000 FA Plus vs TRC-50IX + D7100** | **3D OCT-2000 FA Plus vs TRC-50DX + D7100** |
|---|---|---|---|---|---|---|
| Mean Bifurcation Angle - A | Mean (A / B) | 82.8 / 84.4 | 82.4 / 83.3 | 82.2 / 83.3 | 82.7 / 84.5 | 81.9 / 83.7 |
| | LoA | [-11, 8] | [-11, 9.2] | [-11, 8.4] | [-11, 7] | [-11, 7.1] |
| | Pearson r | .69 (639) | .68 (1534) | .69 (3084) | .72 (4442) | .71 (1276) |
| Median Temporal Angle - A | Mean (A / B) | 124 / 124 | 126 / 126 | 125 / 126 | 126 / 126 | 126 / 126 |
| | LoA | [-26, 27] | [-25, 25] | [-25, 22] | [-23, 23] | [-25, 24] |
| | Pearson r | .70 (547) | .75 (1316) | .76 (2690) | .77 (4384) | .74 (1229) |
| Length-Weighted Caliber - A | Mean (A / B) | 5.92 / 5.92 | 5.76 / 5.76 | 5.79 / 5.79 | 5.98 / 5.98 | 5.92 / 5.92 |
| | LoA | [-.62, .62] | [-.67, .67] | [-.6, .59] | [-.82, .82] | [-.86, .86] |
| | Pearson r | .78 (642) | .76 (1543) | .82 (3087) | .71 (4447) | .68 (1277) |
| Temporal CRE - A | Mean (A / B) | 12.2 / 12.2 | 11.8 / 11.8 | 12 / 12.1 | 12.3 / 12.3 | 12.2 / 12.2 |
| | LoA | [-1.3, 1.3] | [-1.3, 1.3] | [-1.4, 1.3] | [-1.5, 1.5] | [-1.5, 1.5] |
| | Pearson r | .83 (589) | .82 (1396) | .82 (2860) | .80 (4400) | .79 (1258) |
| Full CRE - A | Mean (A / B) | 14.5 / 14.5 | 14.1 / 14.1 | 14.3 / 14.3 | 14.6 / 14.7 | 14.5 / 14.5 |
| | LoA | [-1.3, 1.3] | [-1.3, 1.3] | [-1.4, 1.4] | [-1.6, 1.6] | [-1.6, 1.6] |
| | Pearson r | .87 (586) | .88 (1396) | .85 (2844) | .85 (4377) | .83 (1244) |
| LW Distance Tortuosity - A | Mean (A / B) | 1.02 / 1.03 | 1.02 / 1.03 | 1.02 / 1.03 | 1.02 / 1.03 | 1.02 / 1.03 |
| | LoA | [-.012, .008] | [-.012, .0077] | [-.012, .0072] | [-.013, .0062] | [-.013, .0057] |
| | Pearson r | .93 (642) | .93 (1543) | .94 (3087) | .94 (4447) | .94 (1277) |
| LW Curvature Tortuosity - A | Mean (A / B) | 3.5 / 4 | 3.43 / 3.91 | 3.45 / 4.03 | 3.42 / 4.05 | 3.38 / 4.01 |
| | LoA | [-1.4, .45] | [-1.4, .4] | [-1.5, .36] | [-1.5, .2] | [-1.5, .22] |
| | Pearson r | .90 (642) | .91 (1543) | .92 (3087) | .93 (4447) | .92 (1277) |
| LW Distance Tortuosity [ETDRS] - A | Mean (A / B) | 1.05 / 1.05 | 1.05 / 1.05 | 1.05 / 1.05 | 1.05 / 1.05 | 1.05 / 1.05 |
| | LoA | [-.045, .049] | [-.049, .052] | [-.047, .048] | [-.05, .046] | [-.049, .045] |
| | Pearson r | .67 (543) | .65 (1207) | .71 (2572) | .68 (4050) | .70 (1173) |
| LW Distance Tortuosity [Disc] - A | Mean (A / B) | 1.03 / 1.04 | 1.03 / 1.04 | 1.03 / 1.04 | 1.04 / 1.04 | 1.03 / 1.04 |
| | LoA | [-.018, .014] | [-.017, .013] | [-.016, .013] | [-.017, .013] | [-.017, .013] |
| | Pearson r | .91 (642) | .92 (1543) | .94 (3087) | .93 (4447) | .92 (1277) |
| Temporal Artery-vein ratio | Mean (A / B) | .664 / .664 | .663 / .662 | .674 / .662 | .682 / .663 | .674 / .663 |
| | LoA | [-.088, .087] | [-.097, .099] | [-.073, .098] | [-.071, .11] | [-.087, .11] |
| | Pearson r | .82 (587) | .82 (1390) | .85 (2856) | .83 (4373) | .82 (1255) |
| Nasal Artery-vein ratio | Mean (A / B) | .733 / .742 | .726 / .734 | .738 / .732 | .745 / .723 | .737 / .721 |
| | LoA | [-.12, .1] | [-.13, .11] | [-.1, .12] | [-.092, .14] | [-.1, .14] |
| | Pearson r | .83 (580) | .82 (1378) | .83 (2831) | .81 (4356) | .78 (1239) |
| Full Artery-vein ratio | Mean (A / B) | .644 / .645 | .638 / .64 | .65 / .637 | .657 / .637 | .649 / .636 |
| | LoA | [-.056, .053] | [-.059, .054] | [-.037, .063] | [-.035, .074] | [-.041, .066] |
| | Pearson r | .85 (586) | .84 (1396) | .87 (2843) | .85 (4376) | .84 (1244) |
| Vessel Density [Ellipse] - A | Mean (A / B) | 5.87 / 5.55 | 5.5 / 5.33 | 5.61 / 5.62 | 5.64 / 5.88 | 5.63 / 5.94 |
| | LoA | [-.71, 1.4] | [-.8, 1.1] | [-1.1, 1.1] | [-1.2, .73] | [-1.2, .6] |
| | Pearson r | .70 (642) | .72 (1543) | .68 (3087) | .76 (4447) | .77 (1277) |

## Biomarker sensitivity analysis

Lower image quality levels are likely to degrade the quality of input segmentations used by retinalysis-vascx. In this analysis, we evaluate how different biomarkers perform across image quality levels. Because the degradation of segmentations due to low quality may be difficult to simulate, we instead perturbed source images with domain-relevant perturbations of different strengths before extracting biomarkers. We characterize the change in biomarker values (MAE and ICC(2,1) relative to the biomarker values on the original non-perturbed image with respect to the strength of the perturbation applied. In this way, we provide a measure of biomarker sensitivity to image degradation. A core feature of VascX is its ability to output a null value when the biomarker is not computable on a given image due to a lack of input data or because the computation regions are out of bounds. Next to MAE and ICC metrics, we characterize how the frequency of non-computable outputs changes with perturbation strength.

Our perturbation pipeline applies artifacts and other image transforms designed to mimic degradations present in real fundus images. The pipeline simulates, in order, difuse artifacts (vitreous_floaters, lens_dust), uneven_illumination (dark regions, vignetting), chromatic aberration, and blur (defocus blur, motion blur), and sensor noise. Supplementary Appendix C shows a sample figure with augmented versions of four images.

Figures 7, 8, and 9 show the results of our biomarker sensitivity analysis for a representative set of biomarkers. We normalized MAE relative to the standard deviation of the original biomarker (non-perturbed) to obtain a measure that is comparable across biomarkers. ICCs were calculated following the same ICC(2,1) procedure used in the previous section.

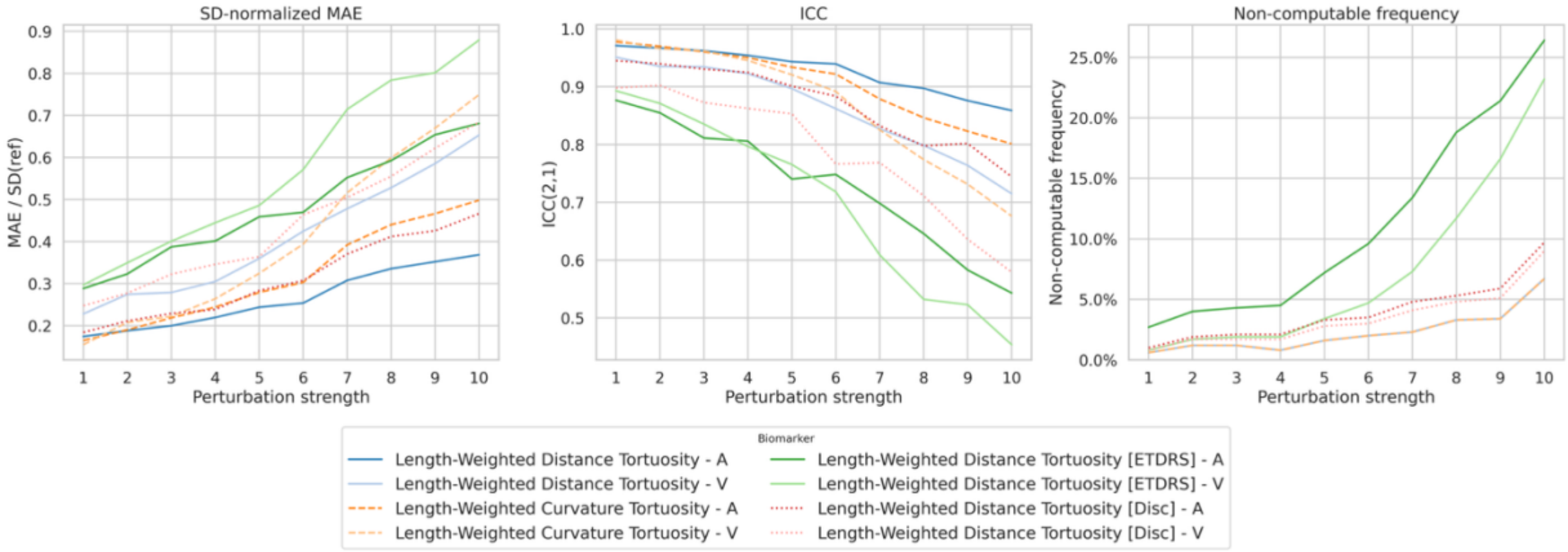

*Figure7. Results of our biomarker sensitivity analysis for tortuosity biomarkers. The plot shows normalized MAE and ICC relative to the non-perturbed biomarker values (Y axes) against the strength of the perturbation applied to the image (X axis). The frequency of non-computable or null outputs per biomarker is shown on the right.*

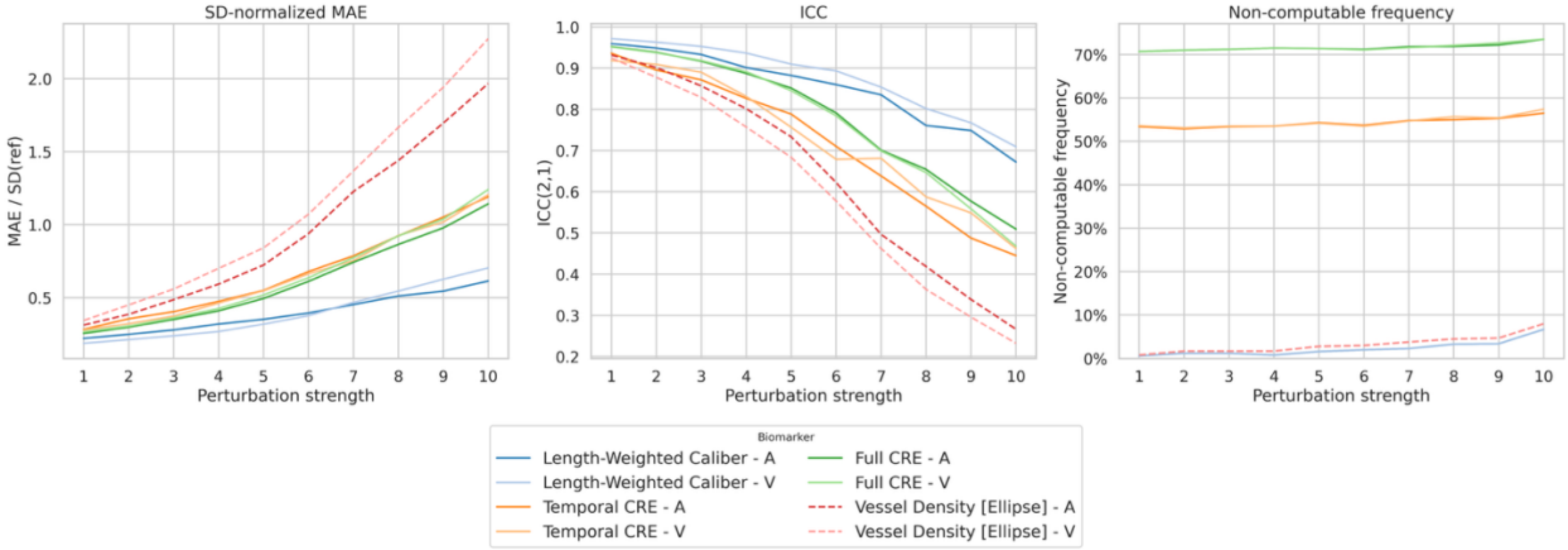

*Figure8. Results of our biomarker sensitivity analysis for caliber, CRE, and vascular density biomarkers. The plot shows normalized MAE and ICC relative to the non-perturbed biomarker values (Y axes) against the strength of the perturbation applied to the image (X axis). The frequency of non-computable or null outputs per biomarker is shown on the right.*

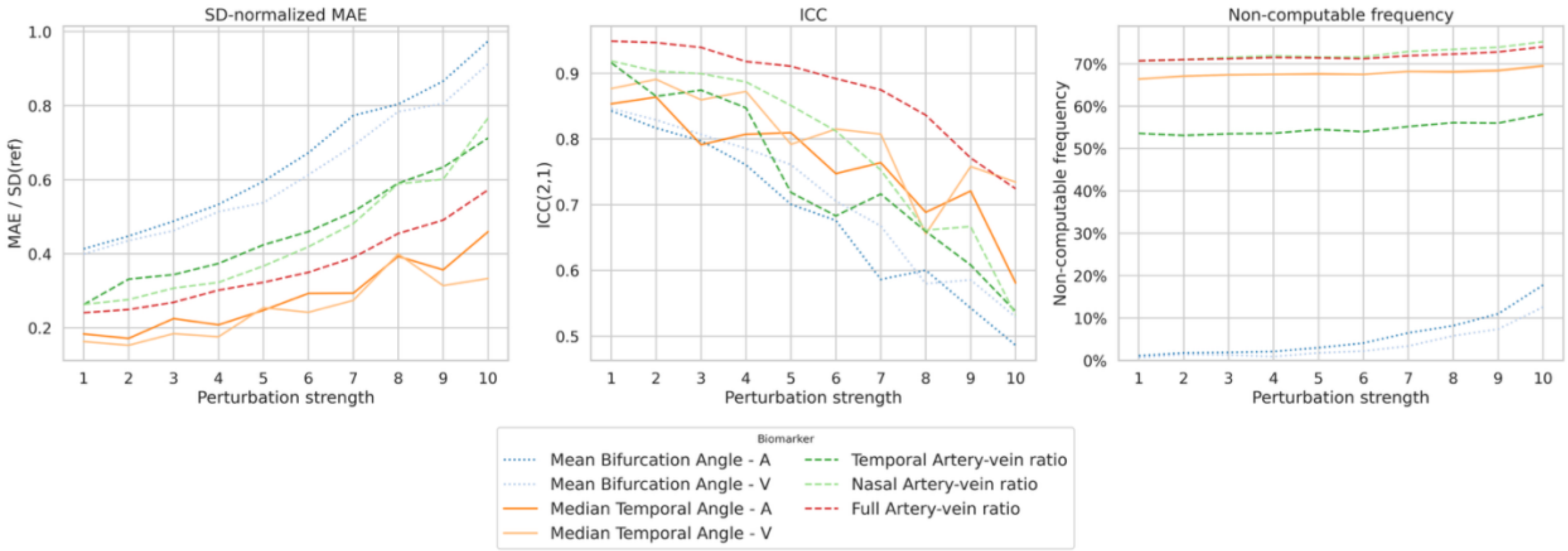

*Figure9. Results of our biomarker sensitivity analysis for bifurcation angle, temporal angles, and artery-vein ratio biomarkers. The plot shows normalized MAE and ICC relative to the non-perturbed biomarker values (Y axes) against the strength of the perturbation applied to the image (X axis). The frequency of non-computable or null outputs per biomarker is shown on the right.*

In addition to image quality and device effects, each VascX biomarker has optional arguments that can be used to configure it (Supplementary Appendix A). We complete our sensitivity analysis with experiments on the effect of biomarker arguments on VascX outputs.

We assessed how changes in vascx parameters affect its outputs on a subset of 100 randomly selected images (with QC as the only exclusion criterion) from the Rotterdam Study. For each of the representative biomarkers evaluated in the manuscript, one continuous parameter was changed at a time from its default value to study its effect in isolation while keeping the others constant. Biomarker values were changed in the range (-50%, +50%) relative to their default values. Figure 10 plots the results on the biomarker scale (Y axis).

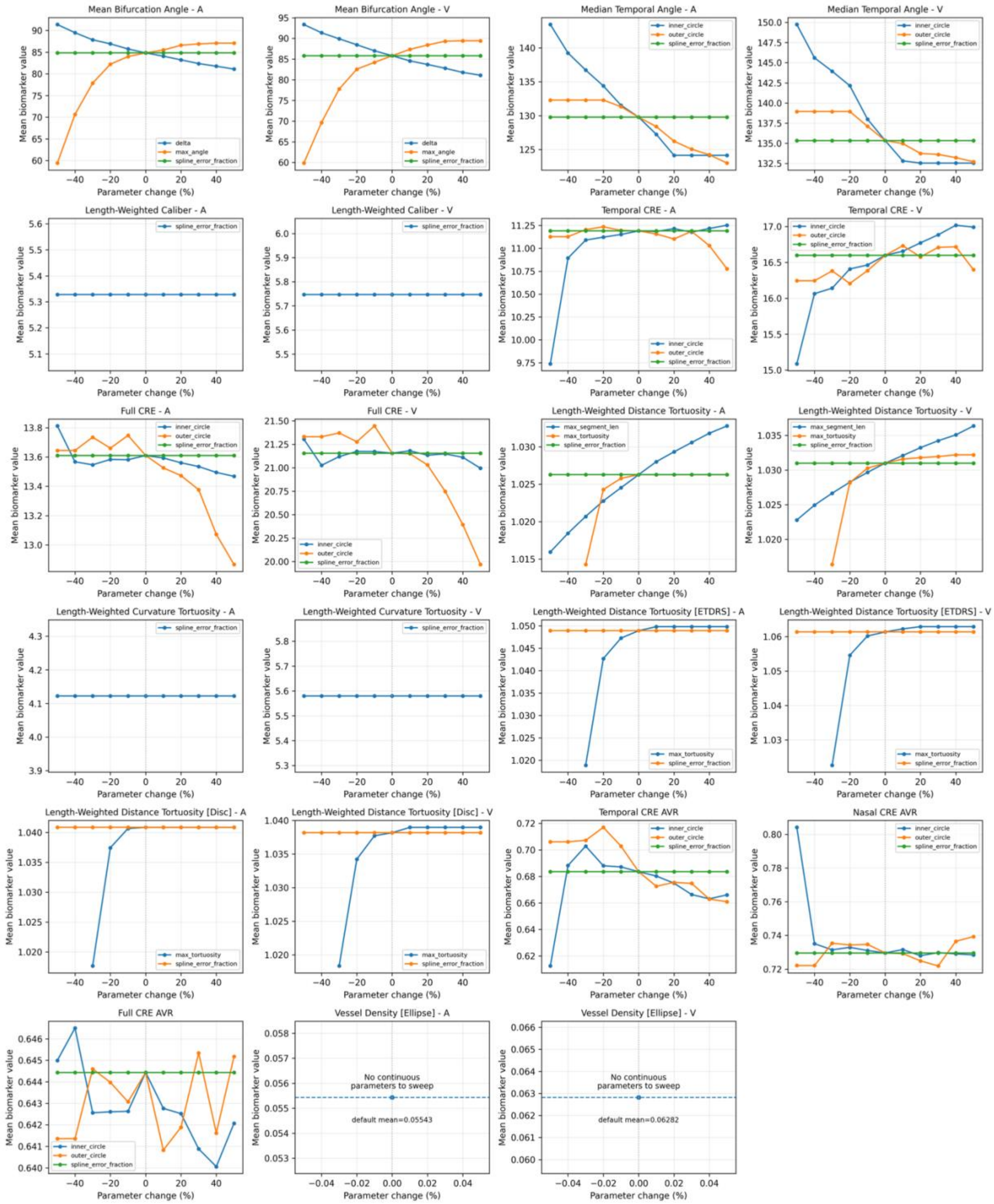

*Figure10. Parameter sensitivity analysis of VascX biomarkers. For each biomarker, the plot shows how changes in the value of continuous heuristic arguments affect the mean value of the biomarker. Plotted points are averages over a sample of 100 images.*

## External usability

Throughout its development, retinalysis-vascx has been used by the authors and external collaborators to run large-scale analysis of the vasculature on both large-scale population data and smaller disease-specific datasets [21–24]. Feedback from the author team and external users has been incorporated into the platform, both to refine the quality of the biomarkers and improve the ease of use of the software. Separate work used VascX as a starting point for a software package designed to measure OCTA vascular biomarkers[25].

# Discussion

We presented and evaluated retinalysis-vascx, a new software toolbox designed to measure retinal vascular biomarkers in an explainable way, using segmentation as an intermediate step. As a main advantage with respect to other software, VascX uses the optic disc segmentation, the fovea location, and the CFI bounds to extract localized biomarkers and to determine when computation regions are out of bounds. At the same time, a wide set of biomarkers has been implemented in base VascX, with some of them – such as tortuosity – having multiple implemented definitions. A representative biomarker set has been evaluated in the manuscript, showing that the platform can extract reproducible biomarkers across a range of devices and image conditions. Biomarker reproducibility, - measured via ICC scores, limits of agreement, and correlation coefficients – and robustness to image perturbations are different per biomarker and measurement region – in line with previous work on same-device reproducibility[14]. Heuristic biomarker parameters have been exposed and characterized to enable systematic and reproducible research on retinal vascular biomarkers.

More specifically, our test-retest reproducibility analysis using extensive CFI imaging from the Rotterdam Study to compare biomarkers on images of the same eye by different imaging devices revealed moderate to good agreement (ICC > 0.5) on most biomarkers and device pairs, with some showing excellent agreement (ICC > 0.9). Bifurcation angles, vascular densities, and tortuosity measured in the ETDRS grid region were the least reproducible biomarkers, with some device pairs showing poor agreement. For bifurcation angles, this can be explained by a lack of robustness against segmentation mistakes such as missing vessels or vessel mislabeling, which can cause a bifurcation to be missed entirely. Vascular density is likely affected by differences in the region imaged. Although we stratified images by field (macula-centered vs OD-centered), different devices had slightly different fields of view and inconsistent centering. Tortuosity in the ETDRS region is likely affected by the lack of visibility of the small vessels in the macular region, which affects segmentation.

On the other hand, whole-image length-weighted tortuosity and CREs stood out as the most reproducible biomarkers. In the case of tortuosity, this could be explained by increased tortuosity manifesting across the entire retinal vasculature, making the biomarker particularly robust. Tortuosity measured on the ETDRS region was considerably less reproducible, likely due to inconsistent visibility of the smaller vessels of the foveal region between devices. CREs, on the other hand, are driven by the caliber measurements of a few, but usually thick vessels close to the optic disc. The thickness of these vessels makes them unlikely to be lost to measurement due to image quality degradation.

Bifurcation angles, calibers, and tortuosity show reduced reproducibility for veins when compared to arteries. This stands out given that vein segmentation is known to have a higher Dice score performance in artery-vein segmentation when compared to arteries[10] and veins, which exhibit higher contrast with the CFI background. In the case of tortuosity, this can be explained by the fact that tortuosity is not invariant to the splitting of a vessel segment (ie., a half circle split in two has lower tortuosity than a connected half circle). For other biomarkers, this trend may be driven by inconsistent visibility of the smallest veins. Small arteries may be consistently under-segmented due to their lower contrast, while small veins are inconsistently segmented, resulting in less consistent vein biomarkers.

Pearson correlations largely mirrored ICC scores, which is expected due to small systematic differences between devices (Table 1). The limits of agreement, indicative of the 95% spread of individual pair differences, varied per biomarker, with most intervals being within $\pm$15% of mean biomarker values, with Length-Weighted Distance Tortuosity (whole image) as narrow as $\pm$1%, CREs and Full Artery-vein ratio close to $\pm$10% on most device pairs, and Median Temporal Angle and Length-Weighted Curvature Tortuosity as high as $\pm$25%. ICC confidence intervals were in most cases relatively narrow, with those of curvature tortuosity being an exception. Inspection of per-device means (Table 1) and the Bland-Altman plots (Supplementary Figure 11) reveals a relatively large difference between devices of around 0.5 for this biomarker, which could explain a wide ICC confidence interval, despite a high ICC, sample size, and correlation. The reason for the systematic difference in curvature is likely related to the computation of the biomarker itself. Although the curvature biomarker is scaled by the optic-disc to fovea distance (used as a proxy for scale) to offset the effect of scale on the curvature equation – curvature is not scale invariant – it is possible that this is an over-correction in practice, or that other factors such as the spline smoothing parameter must be adjusted to maintain scale invariance. Our results suggest that traditional distance-based tortuosity maintains correlation while introducing less systematic bias between devices.

Previous work on same-device reproducibility of AutoMorph is partially in line with our results. Reported ICCs and Spearman *r* are in many cases higher – as expected from the use of the same device and capture conditions between measurements. This is the case for central retinal equivalents (CREs) and Artery-vein ratios (AVRs), with reported ICCs up to 0.95. However, our measured agreement scores for Tortuosity tend to be higher and more consistent across measurement regions. Vascular density scores are similar in magnitude and vary depending on the vessel type measured (arteries or veins). Relative differences between biomarkers are largely consistent with our results, except for tortuosity biomarkers, which we found to be the most reliable.

Our biomarker sensitivity analysis, designed to test robustness against perturbations of the CFIs, showed generally consistent results on both normalized MAE and ICC metrics. As expected, biomarkers exhibited a steady decline with perturbation strength. The lower robustness of bifurcation angles, vessel densities, and ETDRS-localized biomarkers was consistent with the findings of the reproducibility analysis. Tortuosity measurements were, in general, more robust than caliber measurements. Vessel densities were remarkably affected, likely due to the perturbations causing under-segmentation. Global tortuosity, calibers, CRE, and artery-vein ratios were more robust to perturbation.

Interestingly, CREs were less robust to perturbation than length-weighted caliber measurements, contrary to the cross-device reproducibility results in which CREs were slightly more reproducible. This suggests the presence of a different mechanism driving the device differences that is not modeled in the perturbation pipeline. Differences in imaged region or CFI field of view, for example, may be a primary factor driving cross-device differences in length-weighted calibers, as regions away from the optic disc have thinner vessels. This source of error is not present in the image perturbation experiments.

The frequency of null or non-computable outputs (Figures 7-9) showed the expected behavior, with a mild increase from baseline frequencies for most biomarkers. Baseline null frequencies were the highest (~70%) for CREs and temporal angles. This is expected as CREs can only be computed on OD-centered images, which were about 30% of our evaluation set. Temporal Angles similarly require a large region close to the optic disc to be visible in the image, and the OD was too close to the image bounds in many of our macula-centered images. Artery-vein ratios had similar rates due to being derived from CREs. The rest of the biomarkers were computable on most images at baseline, but the frequency of null outputs increased sharply with perturbation strength. The ETDRS / macular region tortuosity biomarkers were the most affected because noise and blur perturbations affect the segmentation of small vessels the most. It must be noted that VascX's ability to output null values in certain non-computable scenarios is not a substitute for a quality control step, which is not part of biomarker computation.

Finally, our parameter sensitivity analysis (Figure 10) showed that most vascx biomarkers exhibit, in most cases, little systematic bias with respect to their optional arguments. The spline smoothing parameter, which controls the smoothness of the splines used to model vessel centerlines, for example, has negligible influence on computation. Some parameters defining computation regions and thresholds, such as inner and outer circles used for Temporal Angle and CRE computation, or the threshold used to filter out invalid tortuosity measurements, however, have a larger effect on the outputs, as expected.

## Potential uses for the software

The VascX pipeline was designed primarily to enable data-driven research and experimentation on retinal vascular biomarkers.

Its main use cases are in oculomics and cardiovascular risk modeling research, where the pipeline can be used to measure standardized, OD–fovea–aligned, and ETDRS-grid biomarkers from CFI images to support association studies and risk prediction for systemic outcomes; potentially helping establish and more deeply understand the associations found in oculomics, or uncovering new ones. Similarly, in clinical and epidemiological research, region-aware measurements (e.g., CRE, temporal arcade angle, tortuosity) in harmonized subfields may help support reproducible, cross-center analyses and longitudinal studies on large biobanks.

The modular, graph-based pipeline (skeleton → undirected graph → directed tree → resolved vessels) was designed to simplify prototyping and comparison between new biomarkers and biomarker variants and aggregation strategies, supporting a deeper understanding of the nuances involved in biomarker computation. Finally, the same modular structure and clear building blocks may help support the adaptation of the platform to different modalities, or its extension with new functionality.

## Limitations of the software

VascX relies on external segmentations (vessels, artery/vein, optic disc, fovea), and their accuracy directly conditions all derived biomarkers. Errors such as missed or spurious vessels, artery–vein misclassification, or mislocalized disc/fovea—often caused by fundus image quality issues (e.g., blur, artifacts, illumination, limited field-of-view)—can bias both global and region-aware (ETDRS, OD–fovea–aligned) measurements. Users should treat segmentation quality and image quality as potential confounders in downstream analyses. Quality control of input images and/or segmentations is highly recommended. Sensitivity analyses or manual validation of biomarker visualizations may be warranted before applying the software. Furthermore, VascX has been developed using model outputs from a specific set of artery-vein segmentation, optic disc segmentation, and fovea localization models developed by the authors. The pipeline's robustness may not be the same if used with inputs from another set of models with different output characteristics.

A second limitation of the pipeline is that it is currently designed specifically for enface images, usually containing either the fovea or optic disc. However, its modular design and clear structure enable its adaptation to other use cases.

# Conclusions

We presented and evaluated retinalysis-vascx, a new software toolbox designed to measure retinal vascular biomarkers in an explainable way, using segmentation as an intermediate step. Our contributions are the following:

- We presented a toolbox that can compute a comprehensive catalog of morphology, topology, caliber, and spatially localized biomarkers from artery–vein segmentations, over regions defined relative to fovea and optic disc landmarks.

VascX enables rapid biomarker prototyping and experimentation via a clear, modular graph-based pipeline and well-documented code.

- We provided a comprehensive characterization of VascX biomarker reproducibility and sensitivity. We present results on the test-retest reproducibility of VascX biomarkers across images of the same eye but captured using different devices. Our results show that most VascX biomarkers have moderate to high test-retest agreement, but with relatively large differences in robustness between biomarkers, which can mostly be explained by the computation procedure. We complemented these results with a sensitivity analysis using artificial perturbations of input images to test the sensitivity of VascX against image quality degradation. These results largely agreed with the test-retest experiments in highlighting different robustness levels across biomarkers. Finally, we characterize how optional parameters of VascX that define computation regions and thresholds affect biomarker outputs, showing that most heuristic parameters have a relatively small impact on biomarker averages.
- The VascX toolbox is open source and easily accessible in Python via PyPI as retinalysis-vascx with the primary goal of supporting ophthalmic and oculomics research.

# Acknowledgments

**Author Contributions:** J.D.V.Q. conceptualized, designed, and developed retinalysis-vascx, conceived the study, developed the methodology, conducted the evaluation, and drafted the manuscript. M.J.B. and S.O.V. contributed to the first implementation of retinal vascular biomarker formulas and provided feedback on the manuscript. S.B. conceived, acquired funding for, and led the VascX Research Consortium that led to this study. C.C.W.K. conceived, helped acquire funding for, and led the biomarker-driven analysis of the retinal vasculature within the consortium. B.L. contributed to securing funding, helped conceive the broader work package within which retinalysis-vascx was developed, contributed code for fitting splines on vessel segments, and gave feedback on the manuscript. Members of the VascX Research Consortium applied retinalysis-vascx in large-scale analyses of the retinal vasculature and provided valuable user feedback.

The authors also acknowledge the staff of the Eyened Reading Center for their contribution to the development of the VascX models used in the study[10]. This work was funded by the Swiss National Science Foundation grant no. CRSII5 209510. Generative AI was used in the preparation of this manuscript as assistance in writing and proofreading. Google Gemini 2 and 3 were used in the period between July 1, 2025 and January 26, 2026.